# Electromagnetically induced transparency at a chiral exceptional point

Changqing Wang[1†], Xuefeng Jiang[1†], Guangming Zhao[1], Mengzhen Zhang[2,3], Chia Wei Hsu[2,3,4], Bo Peng[1,5], A. Douglas Stone[2,3], Liang Jiang[2,3,6] and Lan Yang[1]*

[1]Department of Electrical and Systems Engineering, Washington University, St Louis, MO, 63130, USA.

[2]Departments of Applied Physics and Physics, Yale University, New Haven, CT, 06520, USA.

[3]Yale Quantum Institute, Yale University, New Haven, CT 06520, USA.

[4]Ming Hsieh Department of Electrical Engineering, University of Southern California, Los Angeles, California 90089, USA

[5]Present address: Globalfoundries, East Fishkill, NY USA.

[6]Present address: Pritzker School of Molecular Engineering, University of Chicago, Chicago, IL 60637, USA.

[†]These authors contributed equally.

*Correspondence to: yang@seas.wustl.edu

**Electromagnetically induced transparency, as a quantum interference effect to eliminate optical absorption in an opaque medium, has found extensive applications in slow light generation, optical storage, frequency conversion, optical quantum memory as well as enhanced nonlinear interactions at the few-photon level in all kinds of systems. Recently, there have been great interests in exceptional points, a spectral singularity that could be reached by tuning various parameters in open systems, to render unusual features to the physical systems, such as optical states with chirality. Here we theoretically and experimentally study transparency and absorption modulated by chiral optical states at exceptional points in an indirectly-coupled resonator system. By tuning one resonator to an exceptional point, transparency or absorption occurs depending on the chirality of the eigenstate. Our results demonstrate a new strategy to manipulate the light flow and the spectra of a photonic resonator system by exploiting a discrete optical state associated with specific chirality at an exceptional point as a unique control bit, which opens up a new horizon of controlling slow light using optical states. Compatible with the idea of state control in quantum gate operation, this strategy hence bridges optical computing and storage.**

Originating from the destructive interference between quantum excitation pathways, electromagnetically induced transparency (EIT) is a notable phenomenon in the optical response of a dielectric medium.[1] The transmission spectrum of a probe light is marked by a narrow transparency window in a broad absorption profile, leading to a sharp phase change and a large group delay within the narrow window accompanied by a reduced group velocity for slow light propagation. EIT and its classical analogues have been demonstrated in gas-phase atomic[2–6], metamaterial/metasurface[7,8], plasmonic[9–12], optical[13–15], optomechanical[16–18] and superconducting systems[19,20]. Among them, all-optical analogues of EIT realized in optical resonant systems, such as metamaterial, plasmonic, photonic crystal and whispering-gallery-mode (WGM) resonators, allow for on-chip integration and room-temperature operation. In these optical systems, resonant modes serve as the atomic levels and are coupled via field overlap or additional optical channels such as waveguides, making the systems free from the additional pump light source usually used in gas-phase atomic EIT configuration. Besides, optical platforms provide a high tunability for EIT via engineering resonant frequencies, coupling strengths, and phase delays, which usually rely on the external control of continuous parameters such as temperature or optical power. However, the external sources used in these methods not only add difficulties to the realization of a fully integrated design, but also bring additional noise and instability due to—*e.g.*, thermal fluctuation and nonequilibrium in thermodynamic process.

Quantum and classical state controls of optical properties could harness light–matter interactions under unique physical principles and have found practical applications in photonic communication and computation, such as qubit logic gate[21,22], optical bit storage[23], slow light[24], *etc*. As the manipulation of optical states has become increasingly practical with the fast-growing cavity-engineering techniques, including field backscattering[25], index/material modulation[26] and geometry deformation[27], it is possible to directly manipulate the resonance spectrum – the discrete quasi-normal modes of the system, to fullfill the demanding conditions of EIT/EIA. For example, recent development of $LiNbO_3$ fabrication technology enables robust refractive index modulation in microdisks via digital electrical signals[28,29]. With such technique, the engineering of the discrete optical modes intrinsiclly existing in photonic systems could promise the robust manipulation of the optical anologues of EIT in the integrated photonic platform. Up to now, those advantages of optical states haven't been utilized as the control elements for all-optical EIT in previous studies, due to the lack of an explicit physical scheme that could build a connection between the manipulation of optical states and EIT.

Our result brings revenue to this issue by building on the recent progress in non-Hermitian optics, which has shed light on exotic optical behavior and drawn new attention to unique features of non-hermitian systems. Non-Hermitian wave equations for open systems yield complex energy spectra with non-orthogonality of eigenstates (quasi-normal modes or resonances), which have been utilized to engineer unconventional optical behavior and functionality.[30,31] Unique phenomena occur at exceptional points (EPs), a degeneracy of the complex spectrum, at which two eigenstates coalescence[32], which have enabled unconventional effects, such as unidirectional reflection/invisibility[33], chirality[25], enhanced sensitivity[34,35], topological optical switching[36,37], *etc*., and been applied to engineering light transport, improved sensing technology and microlasing[26,38]. In whispering-gallery-mode (WGM) microresonators, two routes to reaching EPs have been explored: first, by tuning gain-loss contrast and intermodal coupling strength in coupled resonators[39,40], or second, by precisely manipulating backscattering/refractive index distribution in one resonator[25]. In particular, tuning a WGM resonator to an EP, breaks the chiral symmetry of clockwise (CW) and counterclockwise (CCW) light propagation, leading to asymmetric backscattering between the CW and CCW modes. The single merged eigenstate at the EP can be characterized by its chirality, the ratio of the light circulating in the CW and CCW directions normalized to vary between [-1, 1][25]. When the eigenstate involves only CCW components, the chirality approaches -1, and vice versa. Typical modes of an unperturbed resonator are non-chiral. However modes at EPs are associated with either chirality 1 or -1, naturally providing one bit of information storage that can act as a control bit.

Here, we show, both theoretically and experimentally, a way to control optical analogues of EIT in a system of two indirectly coupled WGM microresonators, by directly constructing chiral optical states with the help of EPs. We consider a system as shown in Fig. 1a, where two WGM microresonators $\mu R_1$ and $\mu R_2$ are coupled to a waveguide with coupling strengths $\gamma_{1,1}$ and $\gamma_{1,2}$, respectively. Resonator $\mu R_1$ ($\mu R_2$) supports WGMs with resonant frequency $\omega_1$ ($\omega_2$) and intrinsic loss rate $\gamma_{0,1}$ ($\gamma_{0,2}$). The clockwise (CW) and counterclockwise (CCW) modes are coupled with each other via backscattering on the resonator surfaces. In $\mu R_2$, we assume the coupling strength of the scattering from CW to CCW ($\kappa_{b12}$) is identical to that from CCW to CW ($\kappa_{b21}$). In $\mu R_1$, asymmetric backscattering can be introduced by additional perturbations, *i.e.,* the scattering-induced coupling from CW to CCW ($\kappa_{a12}$) is not equal to that from CCW to CW ($\kappa_{a21}$). The dynamics of the system can be described by the coupled mode equations, given by:

$$\frac{da_1}{dt} = -i\omega_1 a_1 - \frac{\gamma_{0,1} + \gamma_{1,1}}{2} a_1 - i\kappa_{a21} a_2 - \sqrt{\gamma_{1,1}} a_{in},$$

$$\frac{da_2}{dt} = -i\omega_1 a_2 - \frac{\gamma_{0,1} + \gamma_{1,1}}{2} a_2 - i\kappa_{a12} a_1 - e^{i\theta}\sqrt{\gamma_{1,1}\gamma_{1,2}} b_2,$$

$$\frac{db_1}{dt} = -i\omega_2 b_1 - \frac{\gamma_{0,2} + \gamma_{1,2}}{2} b_1 - e^{i\theta}\sqrt{\gamma_{1,1}\gamma_{1,2}} a_1 - i\kappa_{b21} b_2 - e^{i\theta}\sqrt{\gamma_{1,2}} a_{in},$$

$$\frac{db_2}{dt} = -i\omega_2 b_2 - \frac{\gamma_{0,2} + \gamma_{1,2}}{2} b_2 - i\kappa_{b12} b_1,$$

where $a_1$ ($b_1$) and $a_2$ ($b_2$) correspond to the fields of CW and CCW modes in $\mu R_1$ ($\mu R_2$) respectively. The angle $\theta$ is the phase shift accumulated when light propagates in the fiber between the resonators. We use this theoretical model for numerical simulation in this study.

In our experimental setup, a microtoroid is used as $\mu R_1$ and a microdisk as $\mu R_2$, which are coupled to a fibre taper waveguide. The two resonators support WGMs in the 1447 nm wavelength band with similar polarization; the modes in the microtoid and the microdisk have disparate quality factors ($Q$) of $6.9 \times 10^7$ and $1.6 \times 10^5$, respectively. The system is probed by a laser with scanning frequency which is injected into the left port of the fibre taper. Both the transmission and the reflection signals are monitored by photodetectors with the help of an optical switch and two circulators. We first individually characterize the microtoroid ($\mu R_1$) and microdisk ($\mu R_2$) resonators. In the transmission spectra, the mode splitting is observable for the high-$Q$ modes in the microtoroid (Fig 1b), indicating that the degeneracy of CW and CCW modes is broken through some intrinsic perturbation (e.g. geometry or defect-based). For the low-$Q$ modes in the microdisk, the mode splitting is observable only when they are weakly coupled to the taper (Fig. 1c), because makes it difficult to resolve the separation of the two dips in the transmission spectrum when the coupling is large. For both resonators, the reflection spectra (Fig 1b and 1c) also indicate the existence of backscattering, caused by features such as surface roughness and nano-particle accumulation.

The optical analogues of EIT in resonator systems originate from cancellation of intracavity fields due to the destructive interference of light[41]. In particular, a resonant mode is excited not only by the input light field directly, but also by the modal coupling to another resonant mode. These two excitation pathways can destructively interfere with each other and lead to EIT. In our system, light propagation in both forward and backward directions is allowed in the fibre taper between the two resonators, which could be modeled as forward and backward waveguide channels. With the help from these channels and the backscattering in each resonator, a loop is formed in the optical path (Fig. 1a): Light from mode $a_1$ in μR$_1$ couples to the fibre in forward direction ($\sqrt{\gamma_{1,1}}$); after propagating in the fibre with phase modified by the optical path length ($e^{i\theta}$), it couples into mode $b_1$ ($\sqrt{\gamma_{1,2}}$) in μR$_2$, where the backscattering enable the light to couple to mode $b_2$ ($\kappa_{b12}$); subsequently the light couples to the fibre in backward direction ($\sqrt{\gamma_{1,2}}$), propagating backwards accumulating phase ($e^{i\theta}$); when the light meets μR$_1$, it couples into mode $a_2$ ($\sqrt{\gamma_{1,1}}$); through backscattering from CCW to CW, if there is any, the light couples back to mode $a_1$ ($\kappa_{a21}$). The field travelling a round trip in the loop will accumulate a phase shift $\Delta\phi$ which is equal to the phase angle of the coefficient $\gamma_{1,1}\gamma_{1,2}\kappa_{a21}\kappa_{b12}e^{2i\theta}$. At the same time, mode $a_1$ is directly excited by the forward propagating field in the fibre taper. Therefore, optical interference occurs between these two optical paths for the excitation of mode $a_1$. In addition, mode $b_1$ can also be regarded as the starting and ending point of the optical path loop, which indicates that the same type of interference occurs in the excitation of $b_1$. As a result, the field intensity in $a_1$ and $b_1$ can be enhanced by the constructive interference ($\Delta\phi = 2\pi$) or suppressed by the destructive interference ($\Delta\phi = \pi$), which can be controlled by the phase angle $\theta$.

To see chiral states of $\mu R_1$ on the optical interference, we introduce a nanotip to the vicinity of the microtoroid (Fig. 1a), which perturbs the evanescent fields of the modes and their backscattering-induced modal coupling in an asymmetric way[25,34]. The perturbation skews the eigenstates of $\mu R_1$ and the resonator can be tuned to an EP with a single merged eigenstate of either chirality 1 or -1. We define 1) EP$_-$: eigenmode rotates in CCW direction corresponding to chirality -1. The backscattering from CW to CCW is nonzero, and the backscattering from CCW to CW is zero, *i.e.* $\kappa_{a12} \neq 0$ and $\kappa_{a21} = 0$; 2) EP$_+$: eigenmode rotates in CW direction (chirality +1). The backscattering from CW to CCW is zero, and the backscattering from CCW to CW is nonzero, *i.e.* $\kappa_{a12} = 0$ and $\kappa_{a21} \neq 0$. In each case, the vanishing of $\kappa_{a12}$ or $\kappa_{a21}$ is achieved simply by properly adjusting the radial and azimuthal position of the nanotip near the rim of the microtoroid. At EP$_-$, the coefficient $\gamma_{1,1}\gamma_{1,2}\kappa_{a21}\kappa_{b12}e^{2i\theta}$ becomes 0 and thus the optical path loop is broken (Fig. 1d). As a result, the modes in the two resonators do not interfere with each other and are excited by the waveguide mode sequentially. At EP$_+$, the optical loop path still exists, because $\kappa_{a21} \neq 0$ and hence the coefficient $\gamma_{1,1}\gamma_{1,2}\kappa_{a21}\kappa_{b12}e^{2i\theta}$ does not vanish (Fig. 1e). By proper tuning of the phase angle $\theta$, constructive/destructive loop interference can occur for the excitation of the intracavity fields, which could lead to an absorption or transparency window in the spectrum. Here the transparency (absorption) window is a single narrow peak (dip), as a result from the coalescence of eigenstates at EPs.

To understand the necessary conditions of EIT in this system with EP controlled chirality, we construct energy-level diagrams based on the eigenfrequencies of the resonator modes. Without perturbation by the nanotip, there are two eigenfrequencies in $\mu R_1$ ($\mu R_2$) corresponding to level $\omega_{1,\pm}$ ($\omega_{2,\pm}$) which are coupled to level $0$ with coupling strength $\gamma_{1,1}$ ($\gamma_{1,2}$), where level $0$ represents the level at which the optical modes of resonators are not excited (Fig. 2a). Level $\omega_{1,\pm}$ and $\omega_{2,\pm}$ are coupled indirectly via the taper. With a nanotip steering $\mu R_1$ to the EPs, $\omega_{1,+}$ and $\omega_{1,-}$ will merge and become degenerate. At $EP_-$ with chirality -1, the CCW eigenmode field in $\mu R_1$ cannot be scattered back into CW direction. As a result, the flow can only transfer from level $\omega_{2,\pm}$ to level $\omega_{1,EP_-}$ but not in the opposite way, indicating that these two levels are not coupled (Fig. 2b). On the other hand, at $EP_+$ with chirality 1, CCW field can be scattered back into CW eigenmode, and thus the coupling between level $\omega_{1,EP_+}$ and $\omega_{2,\pm}$ exists (Fig. 2c). If $\gamma_{1,1} \ll \gamma_{1,2}$, we could neglect the coupling between level $\omega_{1,EP_+}$ and $0$, and the system at $EP_+$ will support two sets of $\Lambda$-type levels $\{\omega_{1,EP_+}, \omega_{2,\pm}, 0\}$. In our system with highly dissimilar intrinsic loss rates ($\gamma_{0,1} \approx 0.025\gamma_{0,2}$) for the modes in two resonators, this condition is not hard to satisfy, as long as we choose $\gamma_{1,1} < 40\gamma_{0,1}$ and keep $\gamma_{1,2}$ in the same order with $\gamma_{0,2}$. Therefore, the level diagram at $EP_+$ is analogous to $\Lambda$-type energy levels which conventionally lay the foundation for EIT.

Based on the level diagram at $EP_+$, we can further identify the parameter regimes for which the analogue of EIT could be enabled. First, the coupling between level $\omega_{1,EP_+}$ and level $\omega_{2,\pm}$ needs to be strong in analogy to the requirement of large Rabi frequency of the control laser in atomic EIT configuration. The effective coupling coefficient between level $\omega_{1,EP_+}$ and level $\omega_{2,\pm}$ is proportional to $\gamma_{1,1}\gamma_{1,2}\kappa_{a21}\kappa_{b12}e^{2i\theta}$ and thus can be elevated not only by increasing coupling coefficients $\gamma_{1,1}$ and $\gamma_{1,2}$, but also by enhancing intracavity scattering rates $\kappa_{a21}$ and $\kappa_{b12}$. Therefore, in the experiments, we chose a microdisk with more surface roughness to induce larger backscattering and a microtoroid with a significant reflection signal. Second, the decay rate of level $\omega_{1,EP_+}$ should be much smaller than that of level $\omega_{2,\pm}$ in analogy to the longer lifetime of the metastable state in atomic EIT. This is satisfied by choosing highly dissimilar quality factors for modes in the two resonators that allow $\gamma_{0,1} \ll \gamma_{0,2}$.

In the experiments, we study the transmission spectrum controlled by the two types of chiral eigenstates of $\mu R_1$ associated with $EP_-$ and $EP_+$. We first tune $\mu R_1$ to $EP_-$, which is verified by checking that the reflection signal is zero as the light is injected from the right port (Fig. 3b). The lack of necessary backscattering ($\kappa_{a21} = 0$) eliminates loop interference and gives rise to the exhibition of two overlapping dips with highly different linewidths in the transmission spectrum (Fig. 3a). The interpretation of this phenomenon is similar to the "superscattering" phenomenon where the partial scattering cross sections into uncoupled channels are added without interference[42–44]. The narrow absorption dip assisted by $EP_-$ is also found to be invariant with the phase shift $\theta$ (Fig. 3c), consistent with our theoretical prediction (Fig. 3d). Moreover, the depth of the narrow dip varies with the change of coupling strength between $\mu R_1$ and the fibre taper. In this phase-invariant opaqueness, the energy loss is mainly the absorption loss, together with a minor portion of scattering loss into free space and reflection into the backward channel of the fibre. Therefore, these results verify that the -1 chiral state at $EP_-$ leads to a phase-invariant type of absorption, which we name as exceptional-point-assisted absorption (EPAA).

At $EP_+$, with properly chosen $\theta$ controlled by the distance between resonators, we observe a single narrow peak out of a broad absorption in the transmission spectrum (Fig. 4a, upper panel). The pure transparency lineshape appears only when the coupling strength between $\mu R_1$ and the taper is strong. To demonstrate this, we increase $\gamma_{1,1}$ by reducing the gap between $\mu R_1$ and the taper in 50 nm step, and the narrow dip in the transmission spectrum gradually evolves into a narrow peak (Fig. 4c). Furthermore, to verify that the transparency window is associated with the destructive interference in the optical loop, we control the phase $\theta$ by tuning the horizontal distance between $\mu R_1$ and $\mu R_2$. The height of the peak in the transmission spectrum ($\Delta = 0$) is modified with the variation of the distance (Fig. 4a), a hallmark of interference effect. In theory, when the phase shift is optimized ($\theta = \theta_{opt}$), the peak reaches its highest point, corresponding to destructive interference (Fig. 4b). When the phase shift deviates from the optimized value, the peak goes downward. The deepest absorption dip appears at an angle of $\theta_{opt} + \pi/2$, where the term $\gamma_{11}\gamma_{12}\kappa_{a21}\kappa_{b12}e^{2i\theta}$ is opposite to that under optimized phase, corresponding to constructive interference in the optical loop. In addition, as two eigenstates of $\mu R_1$ coalesce at $EP_+$, the transparency window under destructive interference has a single peak, which we name as exceptional-point-assisted transparency (EPAT).

Incorporating these two types of EPs into the level diagram, we can see that in both cases the levels coalesce and $\Lambda$-type levels are formed. However, the chiral state in $\mu R_1$ can only be effectively coupled to the standing-wave modes in $\mu R_2$ in the case that the chirality is 1.

In another word, the chiral mode has a preferred direction of coupling: it can only be coupled to the mode in its forward propagation path. Moreover, we note that the backward transmission signal (with probing laser injected into the right port) is exactly the same due to the reciprocity of the system, further verifying the direction of coupling associated with chirality is invariant with probing directions.

As we have seen so far, the chirality of states plays an indispensable role in tailoring the light transport and interference in this system. Therefore, chiral states enabled by EPs have great potential in slow and fast light generation and control. In integrated photonic structures, the EPs can be achieved by index modulation or geometry deformation. In particular, the precise engineering of EPs and chiral lasing emission in a single WGM microdisk has been realized based on an InGaAsP integrated platform by periodically arranging Ge and Cr/Ge.[26] With recent progress in fabrication technology of $LiNbO_3$, it becomes increasing promising to utilize electric signals with CMOS-compatible voltages[29] – widely used in integrated electronics platforms – to robustly control the local refractive index in a microdisk and steer it to two types of EPs with chirality -1 and 1. Such electronic approach which is highly compatible with integrated photonic platforms will overcome the thermal noise and nonequilibrium problems in conventional methods, and hence enable state-controlled EIT with high operation rate and fidelity.

In conclusion, we have theoretically and experimentally studied EPAT and EPAA in an indirectly-coupled microresonator system. Discrete optical states are employed as control elements for the occurrence of EPAT and EPAA, with high robustness that is lacked in conventional control methods. This study is expected to motivate further explorations of leveraging optical states with non-Hermitian properties to engineer the optical response of various media or systems, and will benefit the on-chip integrated photonic networks for optical storage and information processing.

**Acknowledgements**

This work was supported by the NSF grant No. EFMA1641109, ARO grant No. W911NF1710189. A.D.S. acknowledges the support of NSF grant No. DMR-1743235. L.J. acknowledges the support of the Packard Foundation (2013-39273).

**Author contributions**

X.J., C.W. and L.Y. conceived the idea and designed the experiments. C.W. and X.J. performed the experiments with help from G.Z. and B.P. C.W. analyzed experimental data with help from X.J. and M.Z. Theoretical background and simulations were provided by C.W. with help from M.Z., C.W.H., L.J. and A.D.S. All authors discussed the results and wrote the manuscript. L.Y. supervised the project.

**Competing interests**

The authors declare no competing interests.

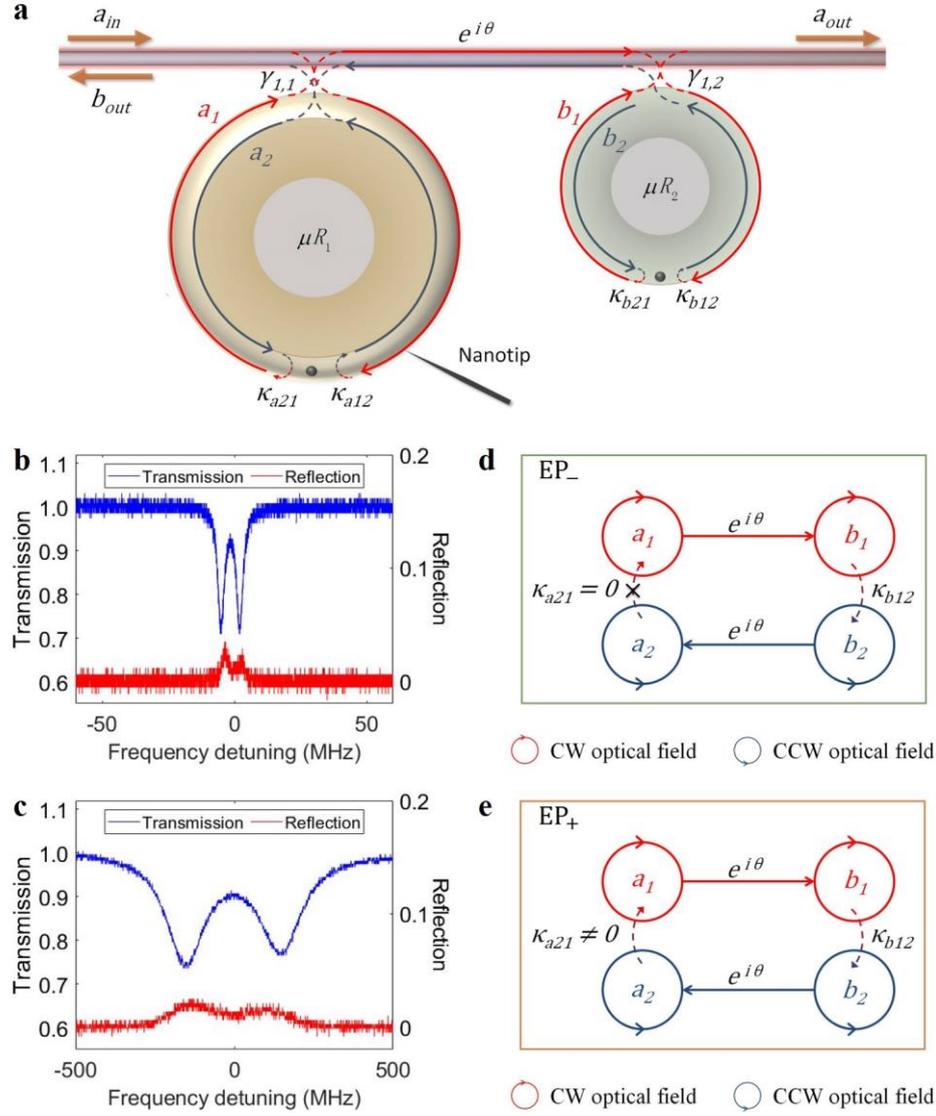

**Fig. 1 | Indirectly coupled WGM microresonators with manipulation of chirality. a**, Schematic diagram of the system consisting of indirectly coupled WGM microresonators. A microtoroid and a microdisk are coupled to a fibre taper with coupling strengths $\gamma_{1,1}$ and $\gamma_{1,2}$ respectively. They both support CW and CCW modes ($a_{1,2}$ and $b_{1,2}$) which have intrinsic loss ($\gamma_{0,1}$ and $\gamma_{0,2}$). Intrinsic perturbations on the surfaces of each resonator can be modeled as an effective scatterer marked by a grey dot, which induces coupling between CW and CCW modes. A nanotip can be applied to the mode volume of $\mu R_1$ to induce asymmetric coupling between the CW and CCW modes and breaks the chiral symmetry of

μR$_1$. An optical path loop is formed: $a_1 \xrightarrow{\sqrt{\gamma_{1,1}}}$ fibre mode ($e^{i\theta}$) $\xrightarrow{\sqrt{\gamma_{1,2}}} b_1 \xrightarrow{\kappa_{b12}} b_2 \xrightarrow{\sqrt{\gamma_{1,2}}}$ fibre mode ($e^{i\theta}$) $\xrightarrow{\sqrt{\gamma_{1,1}}} a_2 \xrightarrow{\kappa_{a12}} a_1$, which interferes with the direct excitation of $a_1$. The same type of interference occurs for $b_1$. The red (blue) curves and lines mark light propagation in the forward (backward) direction. **b-c**, Transmission (blue) and reflection (red) spectra when individually characterizing the modes in μR$_1$ (**b**) and μR$_2$ (**c**) at 1447.6 nm. **d-e**, The optical path loop when μR$_1$ is at EP$_-$ with chirality -1 (**d**) and EP$_+$ with chirality 1 (**e**). CW and CCW optical fields in each resonator are represented by a red circle and a blue circle, respectively. The loop breaks at EP$_-$ because $\kappa_{a21} = 0$, and thus no interference happens. The loop is complete at EP$_+$ because $\kappa_{a21} \neq 0$, and $\theta$ controls the loop interference to be constructive or destructive.

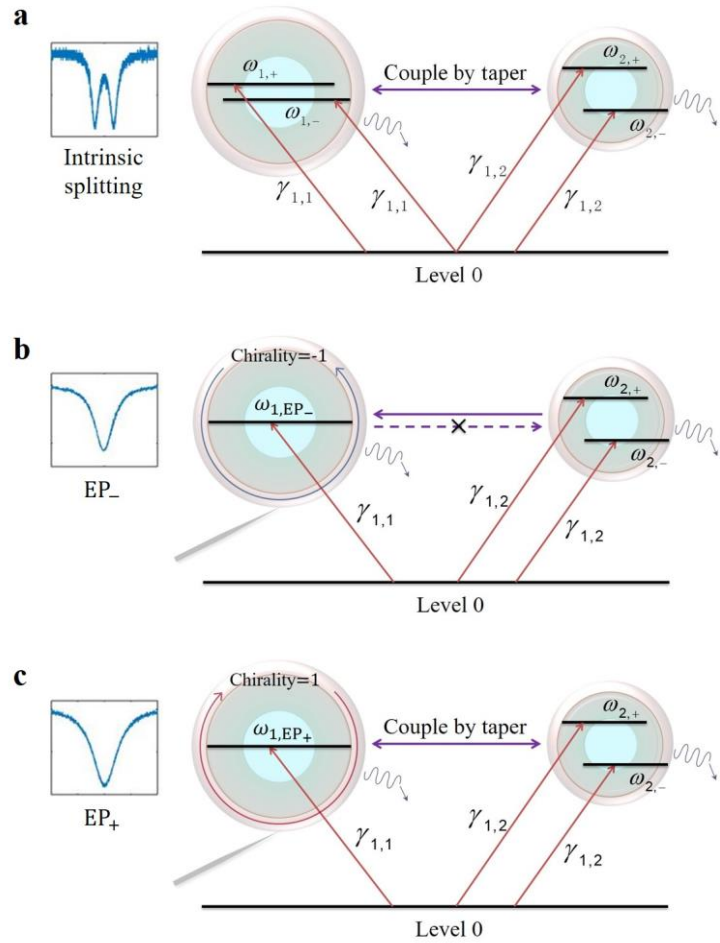

**Fig. 2 | Level diagrams of indirectly coupled WGM microresonators a**, Level diagram of the system when the nanotip is not applied to $\mu R_1$. The degeneracy of each resonator is lifted due to intrinsic perturbations, forming non-chiral standing-wave supermodes. The spectrum of $\mu R_1$ shows mode splitting (inset). Level $\omega_{1,\pm}$ and $\omega_{2,\pm}$ are eigenfrequencies of $\mu R_1$ and $\mu R_2$ respectively, and 0 represents the level at which the optical modes are not excited. **b**, Level diagram of the system when the nanotip steers $\mu R_1$ to $EP_-$ with chirality -1 and CCW eigenmode (blue arc with arrow). The spectrum of $\mu R_1$ shows a single dip (inset). The coupling between two resonators breaks due to lack of backscattering from CCW to CW in $\mu R_1$. **c**, Level diagram when the nanotip steers $\mu R_1$ to $EP_+$ with chirality 1 and CW eigenmode (red arc with arrow). The spectrum of $\mu R_1$ shows a single dip (inset). Due to

$\gamma_{1,1} \ll \gamma_{1,2}$, the level structure can be regarded as two sets of Λ-type levels $\{\omega_{1,\mathrm{EP}_+}, \omega_{2,\pm}, 0\}$.

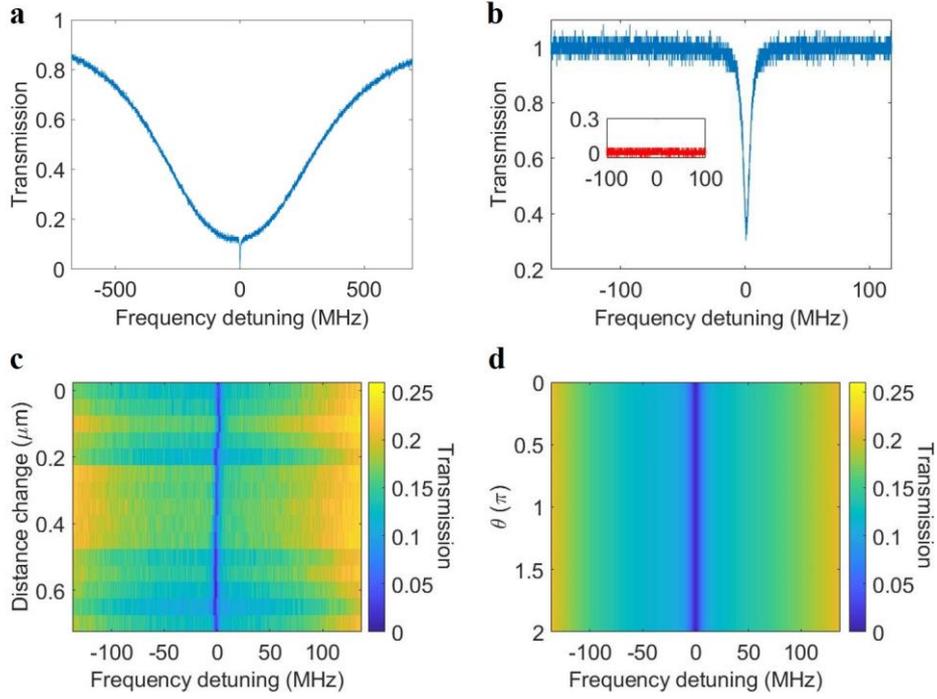

**Fig. 3 | Absorption at EP_ with chirality -1. a**, Experimentally measured transmission spectrum of the system with a randomly picked phase angle $\theta$. **b,** Experimentally measured transmission spectrum when characterizing μR$_1$ with incident light from the right port. The reflection spectrum is flat zero as shown in the inset. **c,** Experimentally measured transmission spectra with the variation of the distance between μR$_1$ and μR$_2$. From top to bottom the distance is changed by a step of 50 nm. **d**, Numerical simulation result of transmission spectra with the variation of $\theta$. Parameters used in the simulation are obtained by fitting the spectrum in **a**: $\kappa_{a21} = 0$, $\kappa_{a12} = (0.2196 - 0.6974i)$ MHz, $\kappa_{b21} = \kappa_{b12} = (0.1327 - 0.0306i)$ GHz, polarization mismatch $\phi = 0.03\pi$.

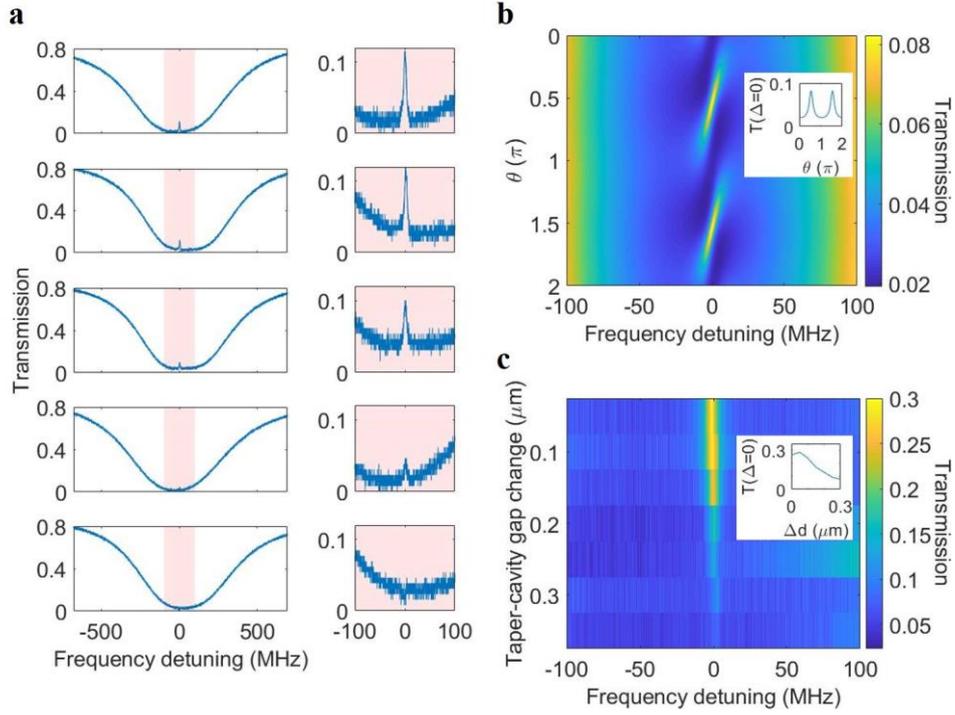

**Fig. 4 | Transparency at $EP_+$ with chirality 1. a**, Experimental measured transmission spectra with different phase angles $\theta$. The close-up transmission spectra within the orange shade region are shown in the insets. $\mu R_2$ is almost critically coupled to the taper and $\mu R_1$ is strongly coupled to the taper. **b**, Numerical simulation result of transmission spectra with the variation of $\theta$. The inset shows the transmission at zero detuning $[T(\Delta = 0)]$ vs. $\theta$. Parameters used in the simulation are obtained by fitting the spectrum in **a**: $\kappa_{a21} = (7.114 - 0.0318i)$ MHz, $\kappa_{a12} = 0$, $\kappa_{b21} = \kappa_{b12} = (0.1337 - 0.0306i)$ GHz, polarization mismatch $\phi = 0.03\pi$. **c**, Experimentally measured transmission spectra with the variation of the gap between $\mu R_1$ and the taper. From top to bottom the gap is increased by a step of 50 nm. The inset shows the transmission at zero detuning $[T(\Delta = 0)]$ vs. the taper-cavity gap change ($\Delta d$).